\documentclass[%
 preprint,
 nofootinbib,
 amsmath,amssymb,
 aps,
 prd,
amsmath,amssymb
]{revtex4-1}

\usepackage{graphicx}
\usepackage{dcolumn}
\usepackage{bm}
\usepackage{hyperref}
\usepackage{multirow}
\usepackage{booktabs}
\usepackage{tabularx}
\usepackage{slashed}
\usepackage{wasysym}
\usepackage[caption=false]{subfig}
\usepackage{xcolor}

\newcolumntype{C}{>{\centering\arraybackslash}X}
\newcolumntype{N}{@{}m{0pt}@{}}

\DeclareMathOperator{\Tr}{Tr}

\begin{document}

\def\be{\begin{equation}}
\def\ee{\end{equation}}
\def\bea{\begin{eqnarray}}
\def\eea{\end{eqnarray}}

\def\tev{\, {\rm TeV}}
\def\gev{\, {\rm GeV}}
\def\mev{\, {\rm MeV}}
\def\kev{\, {\rm keV}}
\def\swsq{\sin^2\theta_W}
\def\xfb{\, {\rm fb}}
\newcommand{\sigmaSI}{\sigma_{\rm SI}}
\newcommand{\sigmaSD}{\sigma_{\rm SD}}
\newcommand{\gsim}{\lower.7ex\hbox{$\;\stackrel{\textstyle>}{\sim}\;$}}
\newcommand{\lsim}{\lower.7ex\hbox{$\;\stackrel{\textstyle<}{\sim}\;$}}
\newcommand{\fb}{\rm fb}
\newcommand{\ifb}{\rm fb^{-1}}
\newcommand{\pb}{\rm pb}
\newcommand{\s}{\rm s}
\newcommand{\yr}{\rm yr}
\newcommand{\cm}{\rm cm}
\newcommand{\sr}{\rm sr}
\newcommand{\kpc}{\rm kpc}
\newcommand{\ipb}{\rm pb^{-1}}

\newcommand{\drawsquare}[2]{\hbox{%
\rule{#2pt}{#1pt}\hskip-#2pt
\rule{#1pt}{#2pt}\hskip-#1pt
\rule[#1pt]{#1pt}{#2pt}}\rule[#1pt]{#2pt}{#2pt}\hskip-#2pt
\rule{#2pt}{#1pt}}

\newcommand{\fund}{\raisebox{-.5pt}{\drawsquare{6.5}{0.4}}}
\newcommand{\Ysymm}{\raisebox{-.5pt}{\drawsquare{6.5}{0.4}}\hskip-0.4pt%
        \raisebox{-.5pt}{\drawsquare{6.5}{0.4}}}
\newcommand{\Yasymm}{\raisebox{-3.5pt}{\drawsquare{6.5}{0.4}}\hskip-6.9pt%
        \raisebox{3pt}{\drawsquare{6.5}{0.4}}}
\newcommand{\antifund}{\overline{\fund}}
\newcommand{\bYasymm}{\overline{\Yasymm}}
\newcommand{\bYsymm}{\overline{\Ysymm}}
\newcommand{\Dsl}[1]{\slash\hskip -0.20 cm #1}

\newcommand{\ssection}[1]{{\em #1.\ }}
\newcommand{\Dsle}[1]{\slash\hskip -0.28 cm #1}
\newcommand{\met}{{\Dsle E_T}}
\newcommand{\Dslp}[1]{\slash\hskip -0.23 cm #1}
\newcommand{\mpt}{{\Dslp p_T}}

\preprint{UH-511-1283-2017}
\preprint{UUITP-30/17}

\title{On the Prospects for Detecting a Net Photon Circular Polarization Produced by Decaying Dark Matter}

\author{Andrey Elagin}
\email{elagin@hep.uchicago.edu}
\affiliation{
Enrico Fermi Institute, University of Chicago, Chicago,  Illinois 60637
}
\author{Jason Kumar}
\email{jkumar@hawaii.edu}
\affiliation{%
 Department of Physics and Astronomy, University of Hawaii, Honolulu, Hawaii 96822
}%

\author{Pearl Sandick}%
 \email{sandick@physics.utah.edu}
\affiliation{%
 Department of Physics and Astronomy, University of Utah, Salt Lake City, Utah 84112
}%
\author{Fei Teng}
 \email{Fei.Teng@utah.edu}
\affiliation{%
 Department of Physics and Astronomy, University of Utah, Salt Lake City, Utah 84112
}%
\affiliation{Department of Physics and Astronomy, Uppsala University, 75108 Uppsala, Sweden }




\date{\today}


\begin{abstract}
If dark matter interactions with Standard Model particles are $CP$-violating, then
dark matter annihilation/decay can produce photons with a net circular polarization.
We consider the prospects for experimentally detecting evidence for such a circular
polarization.  We identify optimal models for dark matter interactions with the
Standard Model, from the point of view of detectability of the net polarization, for
the case of either symmetric or asymmetric dark matter.
We find that, for symmetric dark matter, evidence for net polarization could be found by a search of the
Galactic Center by an instrument sensitive to circular polarization with an efficiency-weighted
exposure of at least $50000~\cm^2~\yr$, provided the systematic detector uncertainties are
constrained at the $1\%$ level.  Better sensitivity can be obtained in the case of
asymmetric dark matter.  We discuss the prospects for achieving the needed level
of performance using possible detector technologies.

\end{abstract}

\maketitle


\section{\label{sec:intro}Introduction}

There are a variety of scenarios in which high-energy astrophysical processes can generate a
gamma-ray spectrum with a net circular polarization (see, for example,~\cite{Ibarra:2016fco,Kumar:2016cum,Gorbunov:2016zxf,Boehm:2017nrl}).
The observation
of such a net circular polarization in the gamma-ray spectrum can thus provide information
about the underlying astrophysics and particle physics, potentially including information
about dark matter interactions.  But one difficulty, noted in previous work, lies in
detecting any net circular polarization.  The goal
of this work is to describe scenarios in which dark matter interactions in astrophysical targets can produce a
circularly polarized photon spectrum which can potentially be observed in future experiments.
We will find that theoretical and experimental considerations will point to a few
scenarios, with optimal detection prospects, which we will use as benchmarks.

A variety of new space-based observatories are contemplated for the ${\cal O}(0.1-100)~\mev$ range,
including ACT~\cite{Boggs:2006mh}, GRIPS~\cite{Greiner:2011ih},
and ASTROGAM~\cite{astrogam}. While these experiments are designed to measure linear polarization and are not sensitive to circular polarization they could provide new constraints on the astrophysical gamma-ray flux in this energy range. For that reason we will refer to these experiments as benchmarks when discussing an efficiency-weighted exposure of an instrument that would be required to measure circular polarization.

Most commonly-used experimental techniques for circular polarization measurements employ Compton scattering, in which the spin and angular dependence of the products of Compton scattering are well-correlated with the initial photon helicity. Therefore it is practical to focus on the photon flux in the energy range $\lesssim {\cal O}(0.3-30)~\mev$ where the Compton scattering cross section is relatively high~\cite{PDG:2016}.

For models of dark matter annihilation or decay that produce a photon spectrum with net circular polarization,
the primary theoretical consideration is the necessity for $P$ and $CP$-violation~\cite{Boehm:2017nrl}.
We will consider two possibilities:
\begin{itemize}
\item{{\it Symmetric dark matter} - The dark matter particle is self-conjugate,
and the initial dark matter state is a $CP$-eigenstate.  In this case dark matter interactions
must violate $CP$, and as a consequence of the optical theorem, a net photon circular polarization can
only be present if dark matter
annihilation/decay also produces a pair of charged particles.}
\item{{\it Asymmetric dark matter} - In this case dark matter interactions need not violate
$CP$, since the initial state itself is not a $CP$-eigenstate.}
\end{itemize}

For symmetric dark matter, these considerations and the requirement that dark matter interactions
survive constraints from Planck~\cite{Ade:2015xua} on
distortions to the Cosmic Microwave Background (CMB) will point to a particular scenario in which dark matter is a
hyperstable (but not exactly stable) spin-0 particle, denoted $X$, which couples to electrons through both scalar and
pseudoscalar interactions.  Dark matter can then decay to two monoenergetic photons ($X \rightarrow \gamma \gamma$)
through a one-loop process in which an electron runs in the loop.  The maximal net circular polarization
which can be achieved in this scenario is $\sim 40\%$ if the mass of the dark matter is $m_X \sim 1.2~\mev$.

However, we will find that, necessarily, the decay process $X \rightarrow e^+ e^-$ can proceed at tree-level,
so the model is thus tightly constrained by bounds on this process from the INTEGRAL experiment~\cite{Jean:2003ci}.
We will see that the best detection prospects arise from a search of the Galactic Center, but even so, would
require an instrument capable of measuring photon circular polarization with $\lesssim 1\%$ systematic uncertainty,
and an efficiency-weighted exposure of $> 50000~\cm^2~\yr$.

On the other hand, if dark matter is asymmetric, then dark matter need not couple to electrons at all.
This scenario is less tightly constrained by other observational constraints, and the experimental
requirements needed for finding evidence of a net circular polarization are relaxed, though still quite
challenging.

The plan of this paper is as follows.  In Section~\ref{sec:GenCon}, we discuss the general theoretical and
experimental considerations which underlie the choice of an optimal scenario for producing a net circular polarization.
In Section~\ref{sec:theory}, we describe the details of the specific models for symmetric and
asymmetric dark matter.
In Section~\ref{sec:Prospects}, we describe the prospects for detecting evidence for these models with a future instrument.  We conclude with a discussion of our results in Section~\ref{sec:Conclusions}.

\section{\label{sec:GenCon} General Considerations}

As noted in~\cite{Boehm:2017nrl}, particle physics processes can only produce a net photon circular polarization
if the processes violate both $P$ and $CP$.  $P$-violation is necessary because the photon flips helicity under
parity.  Moreover, in the absence of $CP$-violation, any net circular polarization generated by
a process would be canceled by the polarization generated by the $CP$-conjugate process.  The necessary
$CP$-violation can arise either from the choice of an initial state, or from dark matter interactions
with the Standard Model.

In the case of asymmetric dark matter, the initial state itself violates $CP$, and in this case it
is quite easy to obtain a circular polarization asymmetry.  For simplicity, we can consider the case
where the asymmetric dark matter, denoted $\chi$, is a hyperstable spin-1/2 particle, which can decay to a left-handed
light fermion $\psi$ ($m_\psi \ll m_\chi$) and a photon\footnote{An example of an interaction term
which could mediate this decay is $(1/\Lambda^2) \bar \psi \sigma^{\mu \nu} P_R \chi F_{\mu \nu}$, where $\Lambda$
is the energy-suppression scale of the effective operator.}.    An example of such a process would be the decay
of a sterile neutrino to an active neutrino and a photon~\cite{Boehm:2017nrl}.
Conservation of angular momentum
necessarily implies that the photon is also left-handed.  The conjugate of the decay process
$\chi \rightarrow \psi \gamma_L$ is the process $\bar \chi \rightarrow \bar \psi \gamma_R$.  The circular polarization
asymmetry is thus determined entirely by the abundances of $\chi$ and $\bar \chi$, and can take any value.

If we instead assume that
the initial dark state is an eigenstate of $CP$, then any $CP$-violation must arise from interactions.
A net circular polarization can then only be generated through one-loop processes, and only if the kinematics
are such that the intermediate particles can go on-shell.  This result follows from the optical theorem.
We can denote by $I$ the initial dark matter state (either a single particle, in the case of a decay process,
or two dark matter particles in the case of dark matter annihilation).
We can similarly denote by $F$ a particular final state, which includes one or more photons.  We denote
by $\bar F$ the $CP$-conjugate final state, in which the helicities of the photons have flipped.  The matrix
elements for the processes $I \rightarrow F$ and $I \rightarrow \bar F$ can be decomposed into $CP$-conserving
($CP$) and $CP$-violating ($CPV$) parts as follows:
\bea
{\cal M}_{I \rightarrow F} &=& {\cal M}_{I \rightarrow F}^{CP} + {\cal M}_{I \rightarrow F}^{CPV} ,
\nonumber\\
{\cal M}_{I \rightarrow \bar F} &=& \pm ({\cal M}_{I \rightarrow F}^{CP} - {\cal M}_{I \rightarrow F}^{CPV}) ,
\eea
where the overall sign of ${\cal M}_{I \rightarrow \bar F}$ above is determined by the eigenvalue of $I$ under
$CP$.

One can only produce a net circular polarization if there is an asymmetry in the rates for the processes
$I \rightarrow F$ and $I \rightarrow \bar F$.  But this asymmetry is in turn proportional to
the difference in the squared matrix elements,
\bea
\Gamma_{I \rightarrow F} - \Gamma_{I \rightarrow \bar F} &\propto& |{\cal M}_{I \rightarrow F}|^2 -
|{\cal M}_{I \rightarrow \bar F}|^2 = 4\,{\rm Re} \left[{\cal M}_{I \rightarrow F}^{CP}({\cal M}_{I \rightarrow F}^{CPV})^* \right] .
\eea

To generate a net circular polarization, it is thus necessary to have both $CP$-conserving and $CP$-violating terms
in the matrix element, and for their relative phase to be different from $\pm \pi/2$.  But for any tree-level process,
the optical theorem guarantees that the $CP$-conserving part of the matrix element will be real, while the $CP$-violating
part will be purely imaginary, ensuring that the asymmetry will vanish.  An acceptable relative phase can only be
generated if the process $I \rightarrow F$ is a loop process, and if the diagram can be cut in such a way that all of the
cut intermediate propagators can go on-shell.

\begin{figure}[t]
\includegraphics[width=\textwidth]{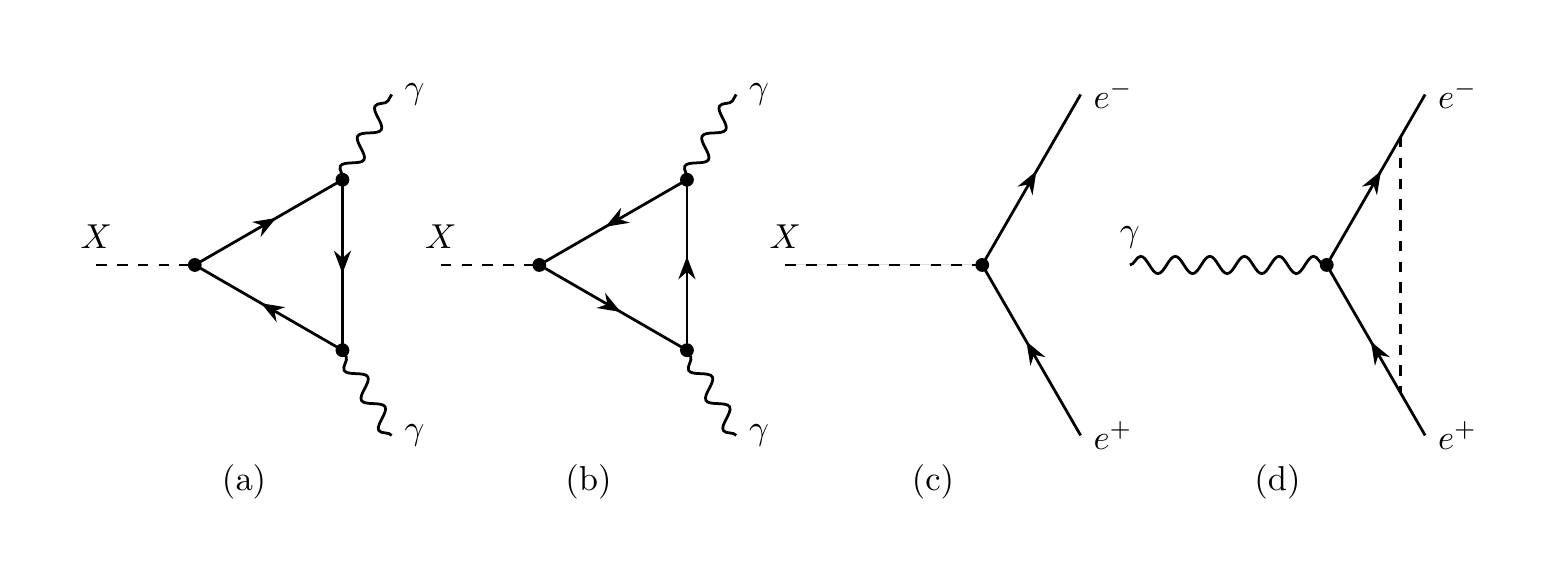}
\caption{\label{fig:feyn} Feynman diagrams of relevant processes:
(a) and (b) one-loop decay of $X$ to photon pairs,
(c) tree level decay of $X$ to $e^+e^-$; (d) vertex correction contributed by $X$ .}
\end{figure}

We have thus found that a net circular polarization cannot be generated from symmetric dark matter
at one-loop if the only final states which are kinematically
accessible have only photons.  Instead, the center-of-mass energy for the process must be at least $\geq 2m_e$, in the case
where the dark matter couples to electrons.  In this case, a photon distribution with net circular polarization
can be produced by the processes $I \rightarrow F, \bar F$ where $F = \gamma^+ \gamma^+$ and $\bar F = \gamma^- \gamma^-$ ($+$ and $-$ denote
the photon helicity), where both processes arise from one-loop diagrams in which an electron runs in the loop [see Figure~\ref{fig:feyn},
panels (a) and (b)].  Of necessity, the process
$I \rightarrow e^+ e^-$ is also kinematically allowed [see Figure~\ref{fig:feyn}, panel (c)].

A similar net polarization could be generated if a muon ran in the loop, provided the center-of-mass energy were greater than $2m_\mu$.  But in
that case, the photons produced from the process $I \rightarrow \gamma \gamma$ would have an energy $\geq m_\mu$.  For photons of this
energy, the dominant detection process is pair production, for which it is difficult to relate any detector observable to the initial
photon helicity.  So although one may construct a variety of models in which $CP$-violating interactions of real dark matter produce photons with a net
circular polarization, this polarization is likely only detectable in the near future if the dark matter couples to electrons, producing
a pair of monoenergetic photons via a one-loop diagram.

Thus far, we have not specified if the process is symmetric dark matter decay or annihilation.  But from the above considerations, we now
see that the best detection prospects lie with dark matter decay ($I = X$).  The reason is because the process $I \rightarrow e^+ e^-$ (which is
necessarily allowed) will occur in the early Universe, and the resulting injection of energetic electromagnetically-coupled particles can distort
the CMB.  The Planck experiment places tight constraints on such CMB distortions~\cite{Ade:2015xua},
in turn constraining the rate for $I \rightarrow e^+ e^-$ in
the early Universe.  But such constraints have the most stringent effect on current observations if the process is dark matter annihilation, because
the rate of dark matter annihilation scales as the square of the dark matter density; the rate for dark matter annihilation was thus much larger in
the early Universe.  Since the rate of dark matter decay scales only as one power of the density, the constraints on the dark matter decay rate arising
from CMB observations are much less stringent.  Although future experiments targeting the ${\cal O}(1-100)~\mev$ range may exceed Planck's sensitivity
to dark matter annihilation with sufficiently large exposure, they should exceed Planck's sensitivity to dark matter decay much more easily~\cite{Boddy:2015efa,Boddy:2015fsa}.

Given this scenario, the only things left to specify are the spin of $X$ and its coupling to electrons.  In order to generate a
net circular polarization, it is necessary for the coupling to have both $CP$-conserving and $CP$-violating terms.
This constraint necessarily forces the choice
of a spin-0 particle.  If the dark matter were instead spin-1, then it could have a renormalizable coupling to
either a vector or axial-vector electron current, but both
choices are $CP$-conserving.

We are thus left with an ``optimal" scenario for the case of either symmetric or asymmetric dark matter:
\begin{itemize}
\item{Symmetric dark matter is a hyperstable real spin-0 particle $X$, which has $m_X > 2m_e$ and which couples to a
linear combination of the
scalar and pseudoscalar electron currents.  The former coupling is $CP$-conserving, while the latter is $P$- and $CP$-violating.}
\item{Asymmetric dark matter is a hyperstable spin-1/2 particle $\chi$ which decays only to a
light left-handed fermion $\psi$ and a photon.}
\end{itemize}

\section{\label{sec:theory} Specific Models}

We begin with the case of symmetric dark matter.
Taking $X$ to be real, the relevant interaction of $X$ with SM particles is given by
\begin{equation}
\label{eq:Lint}
    \mathcal{L}_{\text{int}}=\lambda X\left(\bar{f}P_L f\right)+\lambda^{\ast}X\left(\bar{f}P_R f\right)\,,
\end{equation}
where $P_L$ and $P_R$ are chiral projectors and we take $f=e$. Since the electron mass $m_e$ is kept real, the phase in $\lambda=|\lambda|e^{i\varphi/2}$ cannot be removed by a chiral rotation. Consequently, $\varphi$ leads to $CP$-violation in physical amplitudes that scales as $(m_e/m_X)$.

\subsection{Net photon circular polarization}

In our simplified model \eqref{eq:Lint}, the only prompt photon emission is the one-loop process $X\rightarrow\gamma\gamma$, producing a line spectrum. This process is shown in (a) and (b) of Figure~\ref{fig:feyn}, and the decay amplitudes are
\begin{align}
\label{eq:A2gamma}
    &\mathcal{A}_{\gamma^+\gamma^+}=\frac{2i\alpha_{\text{em}}|\lambda|m_e}{\pi}\left[\left[(1-x)f(x)-1\right]\cos\frac{\varphi}{2}+if(x)\sin\frac{\varphi}{2}\right]\,,\nonumber\\
    &\mathcal{A}_{\gamma^-\gamma^-}=\frac{2i\alpha_{\text{em}}|\lambda|m_e}{\pi}\left[\left[(1-x)f(x)-1\right]\cos\frac{\varphi}{2}-if(x)\sin\frac{\varphi}{2}\right]\,,
\end{align}
where $x \equiv (2m_e/m_X)^2$, the $\gamma^{+,-}$ are photons with positive or negative helicity, respectively, and the function $f$ is
\begin{equation}
\renewcommand{\arraystretch}{1.2}
    f(x)=\left\{\begin{array}{ll}
    -\left[\arcsin(x^{-1/2})\right]^2 & x>1 \\
    \frac{1}{4}\left[\log\frac{1+\sqrt{1-x}}{1-\sqrt{1-x}}-i\pi\right]^2 & x\leqslant 1
    \end{array}\right.\,.
\end{equation}
The derivation of this result is given in Appendix~\ref{sec:decayrate}. When $\varphi=0$ or $2\pi$, we have $\mathcal{A}_{\gamma^+\gamma^+}=\mathcal{A}_{\gamma^-\gamma^-}$ for all $x$, as a consequence of the $CP$ symmetry. For the other values of $\varphi$, we have the following two situations:
\begin{itemize}
    \item If $x>1$, then we have $\mathcal{A}_{\gamma^+\gamma^+}=\mathcal{A}^{\ast}_{\gamma^-\gamma^-}$. This signals a $CP$-violation since $CP$ is a unitary operator. However, the effect of this $CP$-violation only modifies the total decay rate, while the probabilities for producing $\gamma^+\gamma^+$ and $\gamma^-\gamma^-$ final state photons remain the same.  As expected, if $x>1$ then $f(x)$ is real, as a result of the optical theorem, and the net circular polarization vanishes.
    \item If $x\leq 1$, which holds for $m_X \geq 2m_e $, we have $|\mathcal{A}_{\gamma^+\gamma^+}|\neq|\mathcal{A}_{\gamma^-\gamma^-}|$ since $f(x)$ becomes complex. For this case, we do have different decay rates into the two photon final states. Observationally, we will get more photons with one polarization than the other.
\end{itemize}
The total decay rate of $X\rightarrow\gamma\gamma$ is
\begin{equation}
\label{eq:Gamma2g}
\Gamma(X\rightarrow\gamma\gamma)=\frac{1}{2}\times\frac{\left|\mathcal{A}_{\gamma^+\gamma^+}\right|^2+\left|\mathcal{A}_{\gamma^-\gamma^-}\right|^2}{16\pi m_X}=\frac{\alpha_{\text{em}}^2|\lambda|^2m_e}{16\pi^3}\times\left[\sqrt{x}\,\Upsilon(x,\varphi)\right]\,,
\end{equation}
where the explicit factor $1/2$ accounts for the identical final state particles. The function $\Upsilon$ is given by
\begin{align}
    \Upsilon(x,\varphi)=\frac{1}{8}&\,\Bigg\{16\cos^2\frac{\varphi}{2}+\left[(1-x)^2\cos^2\frac{\varphi}{2}+\sin^2\frac{\varphi}{2}\right]\left[\left(\log\frac{1+\sqrt{1-x}}{1-\sqrt{1-x}}\right)^2+\pi^2\right]^2\nonumber\\
    &\quad-8(1-x)\cos^2\frac{\varphi}{2}\left[\left(\log\frac{1+\sqrt{1-x}}{1-\sqrt{1-x}}\right)^2-\pi^2\right]\Bigg\}\,.
\end{align}
The net circular polarization of the final state photons can be defined as:
\begin{equation}
\label{eq:asym}
    R(x,\varphi)=\frac{\left|\Gamma(X\rightarrow\gamma^+\gamma^+)-\Gamma(X\rightarrow\gamma^-\gamma^-)\right|}{\Gamma(X\rightarrow\gamma\gamma)}
    =\frac{\pi |\sin\varphi |}{\Upsilon(x,\varphi)}\log\frac{1+\sqrt{1-x}}{1-\sqrt{1-x}}\,.
\end{equation}
We have compared our analytic expression \eqref{eq:asym} with the numerical results obtained from {\tt FeynArts}~\cite{Hahn:2000kx} and {\tt FormCalc}~\cite{Hahn:1998yk}, and find excellent agreement.

\begin{figure}[t]
\includegraphics[width=0.5\textwidth]{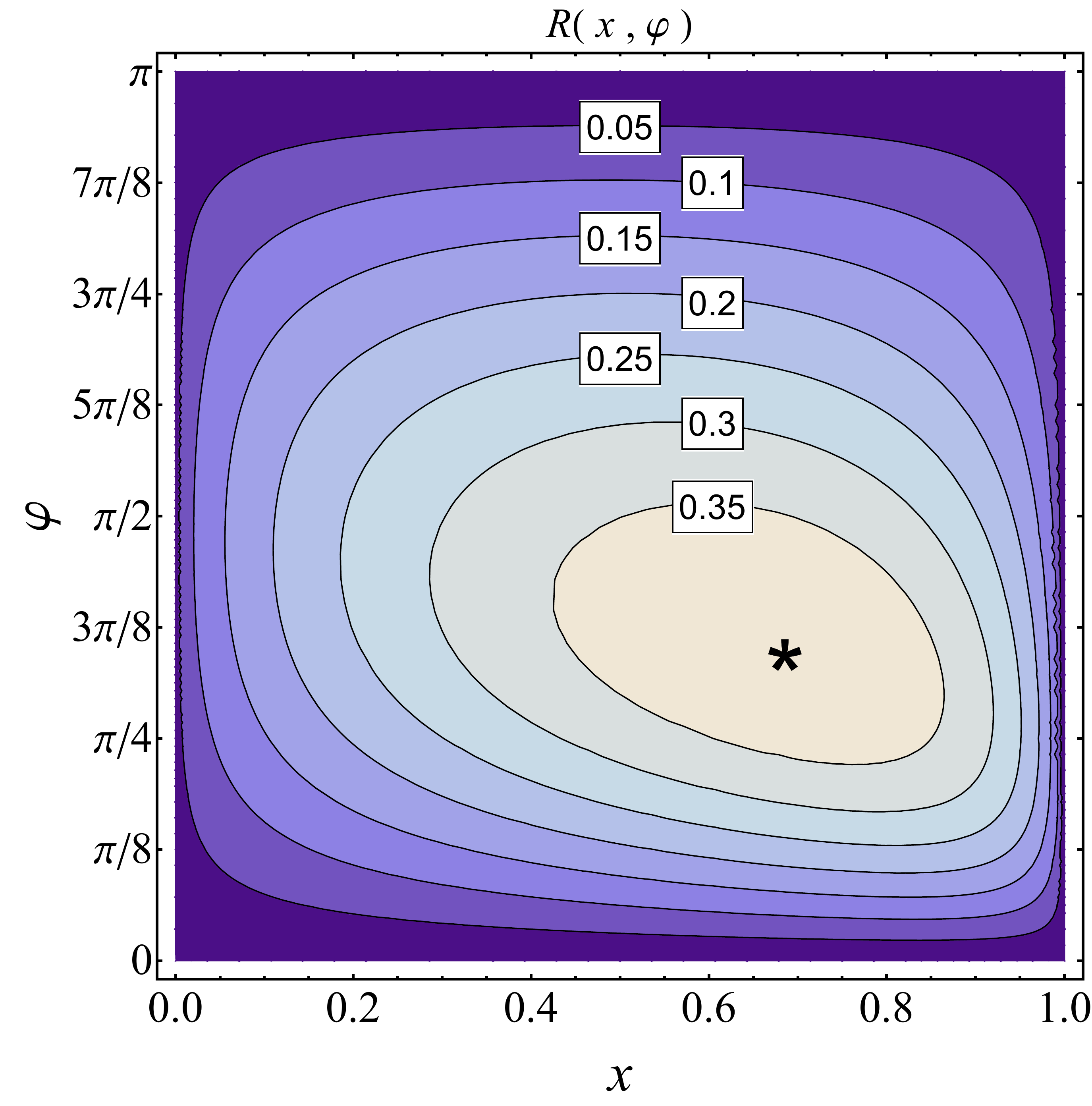}
\caption{\label{fig:x_phi_contour} A contour plot of the net circular polarization $R$
as a function of $x$ and $\varphi$. The star represents the benchmark point as in Table~\ref{tab:bench}.}
\end{figure}

\begin{figure}
\subfloat[]{\includegraphics[width=0.44\textwidth]{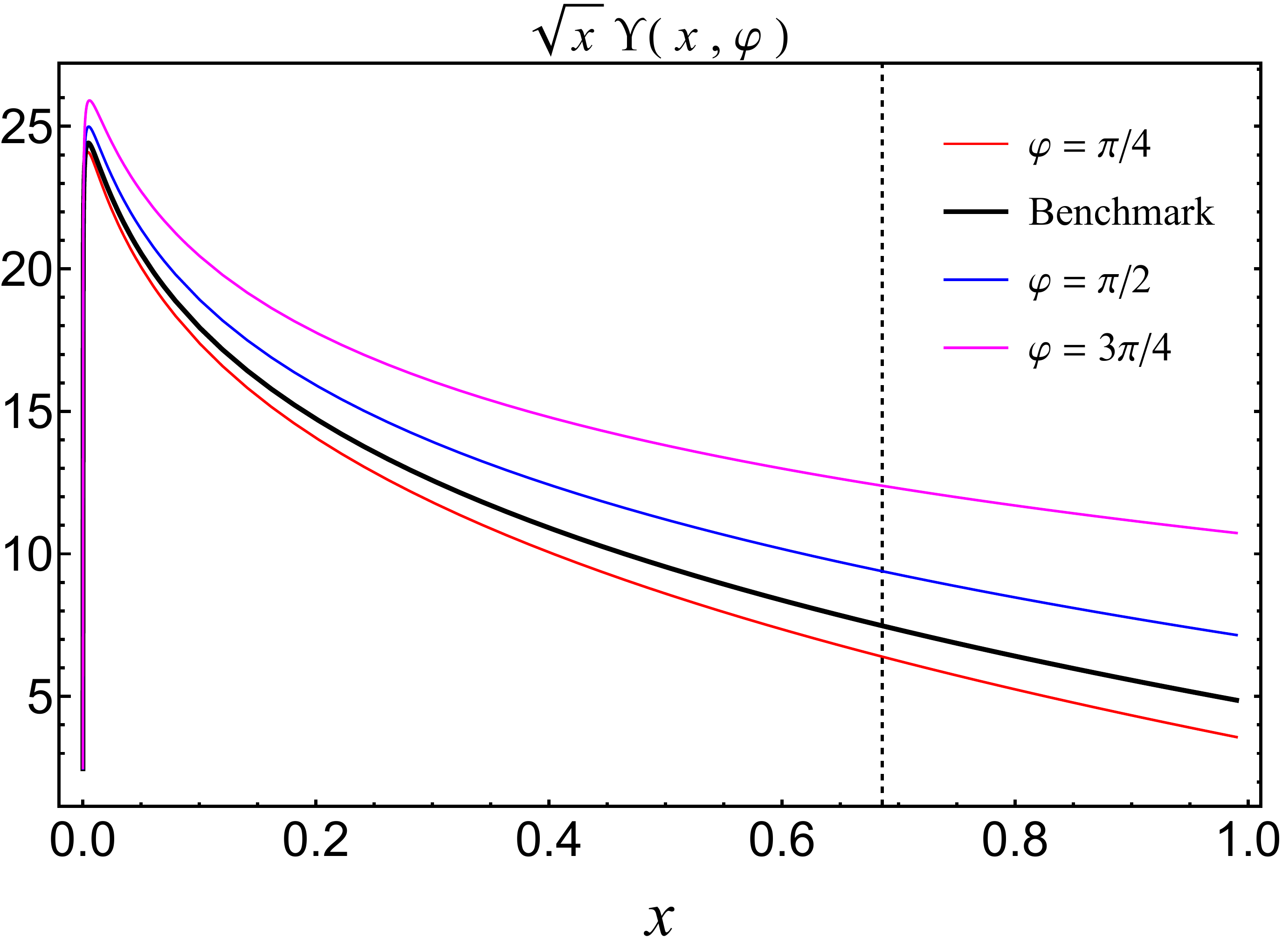}}%
\quad
\subfloat[]{\includegraphics[width=0.44\textwidth]{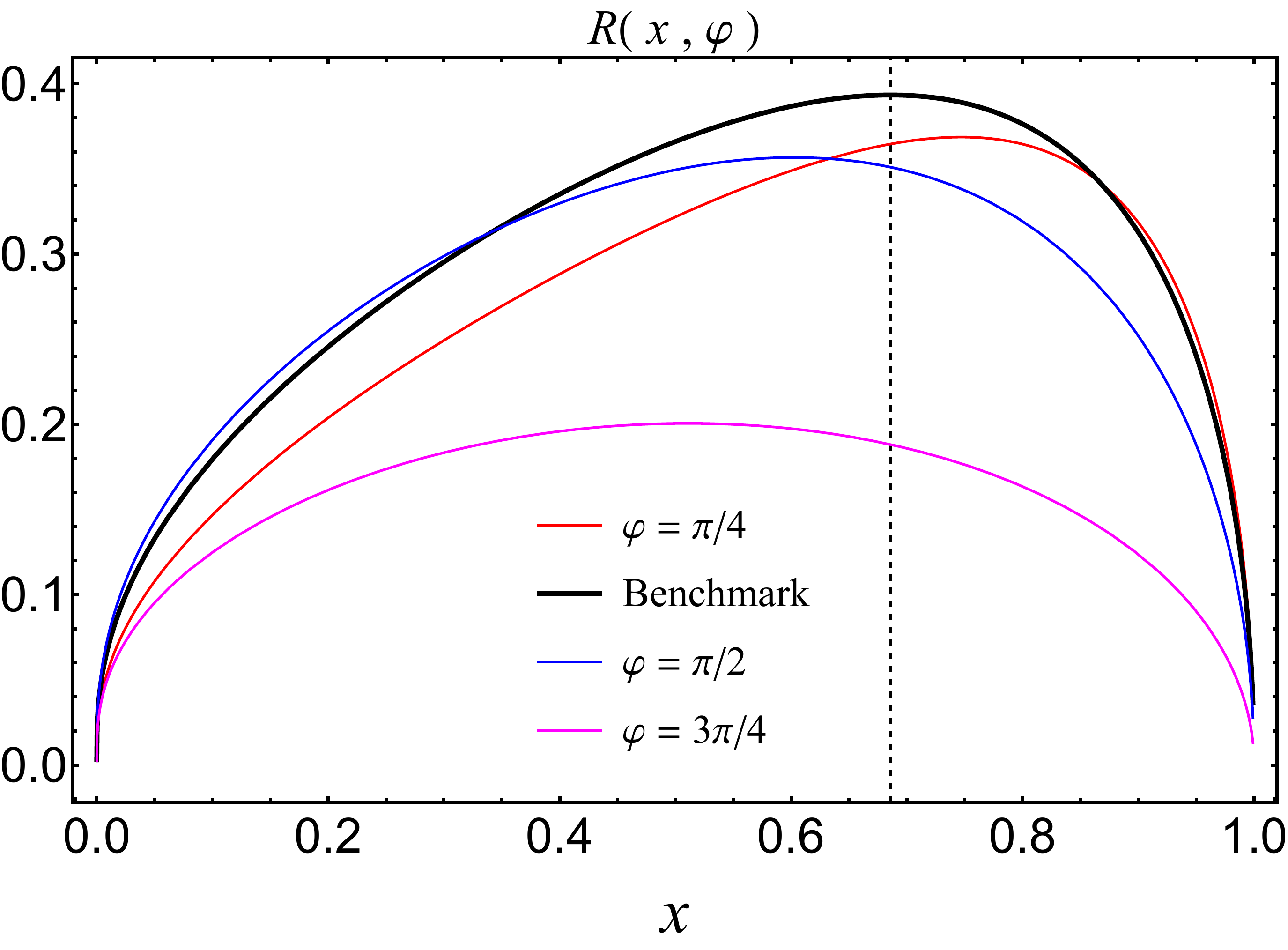}}%
\caption{The decay rate and polarization as a function of $x$. The dashed vertical line represents our benchmark $x$ as shown in Table~\ref{tab:bench}.}
\label{fig:decayrate}
\end{figure}

In Figure~\ref{fig:x_phi_contour}, we present a contour plot of $R$ as a function of $x$ and $\varphi$. The net circular polarization
$R$ is maximized at $R=0.393$ for
$\varphi \sim 1.10$, $x \sim 0.686$ (see Table~\ref{tab:bench}).
In Figure~\ref{fig:decayrate}, we show the dependence of the decay rate, $\Gamma_{\gamma\gamma}\propto\sqrt{x}\,\Upsilon(x,\varphi)$,
and of $R$ on $x$, for a few typical values of $\varphi$. At the boundaries $x=0$ and $x=1$, the function $\Upsilon(x,\varphi)$ behaves as
\begin{align}
    &\Upsilon(x,\varphi)\xrightarrow{x\rightarrow 0}\frac{1}{8}\left(\log\frac{x}{4}\right)^4 & & \Upsilon(x,\varphi)\xrightarrow{x\rightarrow 1}\frac{1}{8}\left[16\cos^2\frac{\varphi}{2}+\pi^4\sin^2\frac{\varphi}{2}\right]\nonumber\\
    &R(x,\varphi)\xrightarrow{x\rightarrow 0}\frac{8\pi|\sin\varphi|}{\left(\log\frac{4}{x}\right)^3} & &R(x,\varphi)\xrightarrow{x\rightarrow 1}\pi\varphi\sqrt{1-x}\,.
\end{align}
Consequently, when we increase $x$ from zero, the decay rate $\Gamma_{\gamma\gamma}\propto\sqrt{x}\,\Upsilon(x,\varphi)$ first increases from zero, reaching its maximum around $x\sim 10^{-3}$, and then decreases. On the other hand, at the branching point $x=1$, the decay rate $\Gamma_{\gamma\gamma}$ increases monotonically with respect to $\varphi$ in the range $0$ to $\pi$. Meanwhile, $R(x,\varphi)$ approaches zero at both $x=0$ and $x=1$.
Note, however, that for $x \lesssim 3 \times 10^{-4}$ we have $m_X \gtrsim 60~\mev$; the resulting photons produced
from the decay of $X$ will likely be too energetic for their circular polarization to be measured.
\begin{table}
\begin{tabularx}{0.75\textwidth}{ |c|c|c|c|C|C|N} \hline
     \multirow{2}{*}{$x$} &  \multirow{2}{*}{$\varphi$} & $m_X$ & \multirow{2}{*}{$R$} & $\Gamma_{X\rightarrow e^+e^-}$ & $\Gamma_{X\rightarrow\gamma\gamma}$ & \\[-3pt]
     &  & (MeV) & & $(\times 10^{-24}\,\textrm{s}^{-1})$ & $(\times 10^{-29}\,\textrm{s}^{-1})$ & \\ \hline
     $0.686$ & $1.10$ & $1.23$ & $0.393$ & $\displaystyle\left(\frac{|\lambda|}{2.18\times 10^{-22}}\right)^2$ & $\displaystyle 2.97\left(\frac{|\lambda|}{2.18\times 10^{-22}}\right)^2$ & \\[2em] \hline
\end{tabularx}
\caption{\label{tab:bench}The benchmark model that maximizes the net circular polarization $R$.}
\end{table}

\subsection{Constraints on \texorpdfstring{$|\lambda|^2$}{|lambda|**2}}

There are a variety of experimental constraints on $|\lambda|^2$, which we detail in this subsection.

\subsubsection{Constraints on \texorpdfstring{$\Gamma(X \rightarrow e^+ e^-)$}{Gamma(x->ee)}}

If $m_X>2m_e$, the main decay channel of $X$ is to electron pairs, with the decay rate:
\begin{equation}
\label{eq:Gammaee}
    \Gamma(X\rightarrow e^+e^-)=\frac{|\lambda|^2 m_X}{8\pi}\left(1-x\right)^{1/2}\left[1-x\cos^2\frac{\varphi}{2}\right]\,.
\end{equation}
The Feynman diagram is shown in (c) of Figure~\ref{fig:feyn}. In order for  $X$ to be approximately stable throughout the
cosmological history, this decay rate must be smaller than the Hubble constant.  This can be translated into
a constraint on $|\lambda|^2$.

On the other hand, a more stringent constraint on the decay rate comes from the CMB. At the era of recombination,
the high energy decay products of dark matter tend to ionize neutral hydrogen atoms. This effect broadens the last scattering surface, which suppresses the high-$\ell$ CMB temperature fluctuation and enhances the low-$\ell$ polarization fluctuation. For $m_X\sim\mathcal{O}(10\,\textrm{MeV})$, we have $\Gamma(X\rightarrow e^+e^-)\lesssim 10^{-24}\,\textrm{s}^{-1}$ from Planck~\cite{Slatyer:2012yq,Ade:2015xua},
which leads to $|\lambda|\lesssim\mathcal{O}(10^{-23})$.

However, an even more stringent constraint arises from observations of the 511 keV line by the INTEGRAL experiment~\cite{Jean:2003ci}.
The low-energy positrons produced when $X$ decays in the Galactic Center will quickly form positronium with
ambient electrons.  The subsequent electron-positron annihilation will produce monoenergetic 511 keV photons.  Current
observations of such photons from the INTEGRAL experiment not only indicate that there is some source of low-energy positrons at the
Galactic Center but that the magnitude of this signal also constrains the magnitude of $\Gamma (X \rightarrow e^+ e^-)$.
It was found in~\cite{Hooper:2004qf} (see also~\cite{Picciotto:2004rp}) that, with a suitable choice of the the dark matter density profile $\rho\sim r^{-0.8}$, the 511 keV excess
seen by INTEGRAL could be fit by a dark matter particle with a decay rate satisfying $\Gamma (X \rightarrow e^+ e^-) \sim 2.5 \times 10^{-27}~{\s^{-1}} \left(\frac{m_X}{\text{ MeV} } \right)$.
However, if we use instead the standard NFW profile~\cite{Navarro:1995iw} with $\rho\sim r^{-1}$ near the Galactic Center, and take into account the uncertainty in the measured flux, this bound can be modified by a factor of several. As a rough guide, we will use
\begin{equation}
    \Gamma (X \rightarrow e^+ e^-) \sim 10^{-26}~{\s^{-1}}
\end{equation}
as our INTEGRAL bound.
If $\Gamma (X \rightarrow e^+ e^-)$ exceeds the above
bound, then the 511 keV photon signal from the Galactic Center would exceed what is observed by INTEGRAL.  For $m_X < 30~\mev$, this
constraint on $\Gamma (X \rightarrow e^+ e^-)$ is always at least an order of magnitude tighter than that arising from
Planck observations.  In our subsequent analysis, we will discuss how our results would change if a more precise bound for the
relevant density profile were determined.

\subsubsection{Constraints on \texorpdfstring{$\Gamma(X \rightarrow \gamma \gamma)$}{Gamma(x->rr)}}

If the total rate for the process $X \rightarrow \gamma \gamma$ were sufficiently
large, then one would observe a monoenergetic photon signal with energy
$E_\gamma = m_X/2$ from the Galactic Center.  The absence of such a signal can be
used to bound $\Gamma (X \rightarrow \gamma \gamma)$, which in turn yields a constraint
on $|\lambda|^2$.

The most conservative constraint one could impose is that the number of monoenergetic
photons expected from dark matter decay in the Galactic Center not exceed the number observed by COMPTEL
in the appropriate energy bin.   Of course, one can obtain a tighter constraint by
modeling the astrophysical background from the Galactic Center.
However, because the decay rate $\Gamma (X \rightarrow \gamma \gamma)$
is suppressed approximately by a factor of $\alpha_{\text{em}}^2(m_e/m_X)$, compared with
$\Gamma (X\rightarrow e^+e^-)$, these constraints are typically subleading compared to that
imposed by INTEGRAL data.

The exception to this rule occurs in the limit $x \rightarrow 1$.  In that case,
since $m_X \rightarrow 2m_e$, $\Gamma (X\rightarrow e^+e^-)$ is phase-space suppressed,
while $\Gamma (X \rightarrow \gamma \gamma)$ is not.  But in the limit $x \rightarrow 1$,
the decay process $X \rightarrow \gamma \gamma$ directly produces two monoenergetic photons with
energy $\sim 511~\kev$.  As such, this process is also constrained by the measurements
of INTEGRAL, yielding a constraint
\begin{equation}
    \Gamma (X \rightarrow \gamma \gamma)_{x \rightarrow 1} \lesssim 10^{-26}~{\s^{-1}}\,.
\end{equation}
This constraint is valid for roughly the range $x > 0.994$, for which the energy of the
monoenergetic photon is within the 3 keV width of the 511 keV line.

\subsubsection{Constraints on dipole moment corrections}

Finally, as shown in (d) of Figure~\ref{fig:feyn}, the Lagrangian \eqref{eq:Lint} also leads to a correction to the anomalous
electric and magnetic dipole moments of the electron~\cite{Fukushima:2013efa}:
\begin{align}
    \Delta a&=\frac{|\lambda|^2\cos\varphi}{4(4\pi)^2}\times\frac{m_e^2\left[(m_e^2-m_X^2)(m_e^2-3m_X^2)+2m_X^4\log\frac{m_e^2}{m_X^2}\right]}{(m_e^2-m_X^2)^3}\,,\nonumber\\
    2m_e\frac{d}{|e|}&=\frac{|\lambda|^2\sin\varphi}{4(4\pi)^2}\times\frac{m_e^2\left[(m_e^2-m_X^2)(m_e^2-3m_X^2)+2m_X^4\log\frac{m_e^2}{m_X^2}\right]}{(m_e^2-m_X^2)^3}\,.
\end{align}
However, the current dipole moment measurements~\cite{Olive:2016xmw} only yield the constraint $|\lambda|\lesssim \mathcal{O}(10^{-7})$, which is far weaker than those arising from CMB observations or from INTEGRAL.

\subsection{Asymmetric Dark Matter}

By contrast, the analysis for the asymmetric dark matter scenario is much simpler.  Since the relevant decay processes
are $\chi \rightarrow \psi \gamma_L$ and $\bar \chi \rightarrow \bar \psi \gamma_R$, the circular polarization $R$ is
determined only by the relative abundances of $\chi$ and $\bar \chi$, and can assume any value between 0 and 1.  The
photon is again monochromatic with an energy $E_\gamma \sim m_\chi /2$ (since we assume $m_\psi \ll m_\chi$).
The only parameter of the model remaining is the total decay width
\begin{equation}
\Gamma_\chi = \Gamma(\chi \rightarrow \psi \gamma_L) +
\Gamma (\bar \chi \rightarrow \bar \psi \gamma_R)\,.
\end{equation}
However, the constraints on the total decay width are much weaker than in the case of symmetric dark matter, since $\chi$ does not couple
to electrons.  In particular, the only relevant constraints are those imposed by Planck measurements of CMB distortions ($\Gamma_\chi \lesssim
10^{-24}~\s^{-1}$), and that the total photon flux arising from dark matter decay not exceed the flux observed by COMPTEL in the
appropriate energy bin.  The latter constraint is dominated by systematic uncertainties, and following~\cite{Bartels:2017dpb}, we will
assume that the fraction of the photon flux  which may arise from dark matter decay in any one energy bin (assuming an energy
resolution of $\sim 2\%$)  must be $\leq 0.02$.

\section{Experimental Prospects \label{sec:Prospects}}

\subsection{Experimental Considerations for the Detection of Circular Polarization of a Gamma-Ray Flux}

In practice, circular polarization of gamma-rays can be measured either through their interactions with polarized particles or through helicity transfer to initially unpolarized particles. Compton scattering of magnetized iron is an example of the former approach~\cite{Schopper:1958zz}. An example of the latter is the measurement of bremsstrahlung asymmetry of secondary electrons produced in Compton scattering of unpolarized matter~\cite{Tashenov:2011}.

If Compton scattering is used as the primary interaction of gamma-rays then a practical energy range of detectable photons is limited to $0.3\sim30~\text{MeV}$~\cite{PDG:2016}. Other detection techniques such as those using the photonuclear effect~\cite{Schopper:1958zz} or pair production~\cite{Schopper:1958zz,Olsen:1959zz,Olsen1962,Kolbenstvedt1965}  are also sensitive to the gamma-ray circular polarization and, in principle, allow the energy range to be extended. However, no efficient methods using non-Compton scattering techniques for gamma-ray circular polarimetry have been developed to date.

All circular polarimetry techniques measure various secondary asymmetries to determine the polarization of the primary gamma-ray flux. Secondary asymmetry is an asymmetry in the detection of particles produced as a result of the gamma-ray flux passage through the detector. Currently available methods require a very high number of incoming gamma-rays to produce one useful event that can be used in the secondary asymmetry measurement.
Expected efficiencies for useful events are $e_p \simeq 10^{-3} \sim 10^{-4}$ for the method involving Compton scattering of magnetized iron~\cite{Schopper:1958zz} and $e_p \simeq 10^{-6}$ for the method using bremsstrahlung asymmetry of Compton scattered electrons of unpolarized matter~\cite{Tashenov:2011}.

Typical asymmetries in the secondary particle spectra expected from a 100\% polarized gamma-ray flux are $A \leq$10\%.
Note that the asymmetry in the secondary particle spectra scales linearly with the incoming gamma-ray flux polarization.
Assuming that polarization of the incoming gamma-ray flux $P_{\gamma}$ and the efficiency $e_p$ are independent of the energy spread of the flux and considering a simple counting experiment one needs a total number of useful events of $N_{\text{useful}}\geq 9\frac{1-(AP_{\gamma})^2}{(AP_{\gamma})^2}$ to observe this flux polarization at the 3$\sigma$ level\footnote{This follows Eq.~\eqref{eq:B7} in Appendix~\ref{sec:appB}, with $F=AP_{\gamma}$.}.
Therefore a total integrated flux of
\bea
N_{\text{total}} = \frac{9}{e_p}\frac{1-(AP_{\gamma})^2}{(AP_{\gamma})^2} \approx \frac{9}{e_p(AP_{\gamma})^2}
\eea
gamma-rays is required to pass through the detector for a $3\sigma$ observation of a flux polarization $P_\gamma$.

\subsection{Detection of Circular Polarization of an Astrophysical Gamma-Ray Flux}
\label{sec:ACTdetector}
We will now consider the experimental prospects for detecting a net circular polarization
in a search of either a dwarf spheroidal galaxy (Draco), or of the Galactic Center.  In
either case, we will take as a benchmark an instrument with roughly the characteristics
of ACT~\cite{Boggs:2006mh}.  That is, we will assume an exposure of $I_{\rm exp} =5000~\cm^2~\yr$
and an energy resolution of $\epsilon =0.01$.  Furthermore, we will assume that the efficiency for useful events
is given by $e_p =10^{-4}$, and that the secondary particle asymmetry
arising from 100\% circularly polarized photons is given by $A=0.1$.
Importantly, we will assume that the astrophysical foregrounds/backgrounds from the direction of the
Galactic Center have no expected net circular polarization; any actual net circular polarization in the
astrophysical backgrounds would thus arise from statistical fluctuations.

The number of useful events arising from astrophysical foregrounds/backgrounds can then be written as
\bea
N_{\text{astro}} &=&  \frac{d^2 \Phi_{\rm astro} (E)}{dE d\Omega} \times \Delta \Omega
\times (2\epsilon E) \times (I_{\rm exp} e_p)\,,
\eea
where $d^2 \Phi_{\rm astro}(E) /dE d\Omega$ is the differential photon flux
arising from astrophysical foregrounds/backgrounds in the direction of the
target, $\Delta \Omega$ is the solid angle over which the target is observed,
and $E = m_e /\sqrt{x}$ is the center of the energy bin which one observes.
The width of the energy bin is $2\epsilon$.

For the case of a dwarf spheroidal galaxy (dSph), since there is relatively little baryonic
matter within the dwarf spheroidal, one expects the predominant foreground/background to be
from diffuse emission due to astrophysical processes along the line of sight.  The diffuse
gamma-ray flux in this energy range has been measured by COMPTEL and EGRET, and can be
roughly approximated by the power law expression~\cite{Boddy:2015efa}
\be
\frac{d^2 \Phi_{\rm astro}^{\rm dSph}(E)}{dE d\Omega} = 2.74 \times 10^{-3} \left(\frac{E}{\mev}\right)^{-2.0} \cm^{-2} \s^{-1} \sr^{-1} \mev^{-1}  .
\ee
Within a box with angular size $60^\circ \times 10^\circ$ ($\Delta \Omega = 0.183$)
at the Galactic Center, the MeV-scale photon background measured
by COMPTEL can be fitted by a power law with an exponential cut-off~\cite{Bartels:2017dpb}\footnote{
This equation appears in the published version of~\cite{Bartels:2017dpb}.  After the publication of this work, 
the authors of~\cite{Bartels:2017dpb} 
confirmed to us that there was a typo in their equation; the energy scale of the exponential cut-off should be at 
$20\mev$, not $2\mev$.  But in the energy range wherein we assume the validity of the fit, the effect of this change 
is small, and does not affect our conclusions.}:
\begin{equation}
\label{eq:GCbackground}
\frac{d\Phi_{\text{astro}}^{\rm GC}(E)}{dEd\Omega}=0.013\left(\frac{E}{\mev}\right)^{-1.8}e^{-\left(\frac{E}{2\mev}\right)^{2}}\cm^{-2}\s^{-1}\sr^{-1}\mev^{-1}\,.
\end{equation}
This fit is only valid for $E$ less than a few MeV.

The number of useful events arising from dark matter decay can be written as
\bea
N_{\rm decay} &=& \frac{\mathcal{J}_{\Delta \Omega}}{4\pi}\,
\frac{\Gamma}{m_X}\times 0.68 \times (I_{\rm exp} e_p)\,,
\eea
where $\mathcal{J}_{\Delta \Omega} \equiv \int_{\Delta \Omega} d\Omega \, d\ell \, \rho (\ell)$
is the $J$-factor for decay which arises from integrating the dark matter density $\rho$ over
the observed solid angle $\Delta \Omega$ and along the line-of-sight.
Here, $\Gamma = 2\Gamma(X \rightarrow \gamma \gamma)$ for the case of decaying symmetric dark matter,
and $\Gamma = \Gamma_\chi = \Gamma(\chi \rightarrow \psi \gamma_L) + \Gamma(\bar \chi \rightarrow \bar \psi \gamma_R)$
for the case of asymmetric dark matter.
The factor of $0.68$ arises
because each photon has a $\sim 68\%$ chance of
being reconstructed within the energy window $[(1 - \epsilon) E, (1 +\epsilon) E]$.  We use the
following benchmarks for observations of Draco~\cite{Geringer-Sameth:2014yza} and of the Galactic Center~\cite{Cirelli:2010xx}:
\bea
\mathcal{J}_{\Delta \Omega}^{\rm Draco} (\Delta \Omega^{\rm Draco} = 0.0016 ) &=&
1.38 \times 10^{18} ~\gev \, \cm^{-2}
\nonumber\\
\mathcal{J}_{\Delta \Omega}^{\rm GC} (\Delta \Omega^{\rm GC} = 0.183) &=&
1.03 \times 10^{22} ~\gev \, \cm^{-2} \,.
\eea
where $\Delta\Omega^{\rm Draco}$ is the angular size of Draco, and $\Delta\Omega^{\rm GC}$ is the angular size of the $60^\circ\times 10^\circ$ box at the Galactic Center.

The total expected number of useful photon events is then given by $N_{\rm total} = N_{\rm astro} + N_{\rm decay}$,
while the expected difference in circularly polarized useful photon events is given by $N_{\rm pol} =R \times N_{\rm decay}$.
Note that both $N_{\rm total}$ and $N_{\rm pol}$ scale linearly with the quantity $(I_{\rm exp} e_p)$.
Moreover, $N_{\rm astro}$ scales linearly with $\epsilon$, while $N_{\rm decay}$ is independent of $\epsilon$.
Using these scaling relations, it will be relatively easy to rescale our results for a different choice of
detector specifications.

For any fixed choice of detector parameters $I_{\rm exp}$, $e_p$, $A$ and $\epsilon$ and for
$E \sim {\cal O}(\mev)$, one finds that ratios $N_{\rm decay}^{\rm GC} / N_{\rm decay}^{\rm Draco} \sim {\cal O}(10^4)$,
while $N_{\rm astro}^{\rm GC} / N_{\rm astro}^{\rm Draco} \sim {\cal O}(10^3)$.  Thus, a search
for dark matter decay in the Galactic Center would not only produce more photon events arising from dark
matter decay than would a search of Draco, but also a larger signal-to-background ratio.
Typically, the major advantage of a dark matter search in a dSph, compared to the Galactic Center, is that
the systematic uncertainties in the astrophysical background are better under control, since
they can be measured by off-axis observation.  However, in a search for net circularly polarized
monoenergetic photons, the systematic uncertainties in astrophysical background from the Galactic Center should also be
under control: as there is no known astrophysical origin of a line signal with net circular
polarization, any asymmetry of astrophysical origin would be expected to arise from statistical fluctuations.
Thus, we will focus on the Galactic Center for the remainder of this analysis.

\subsection{Detection Prospects for a Search of the Galactic Center for Symmetric Dark Matter Decay}

\begin{figure}[t]
\includegraphics*[width=0.49\textwidth]{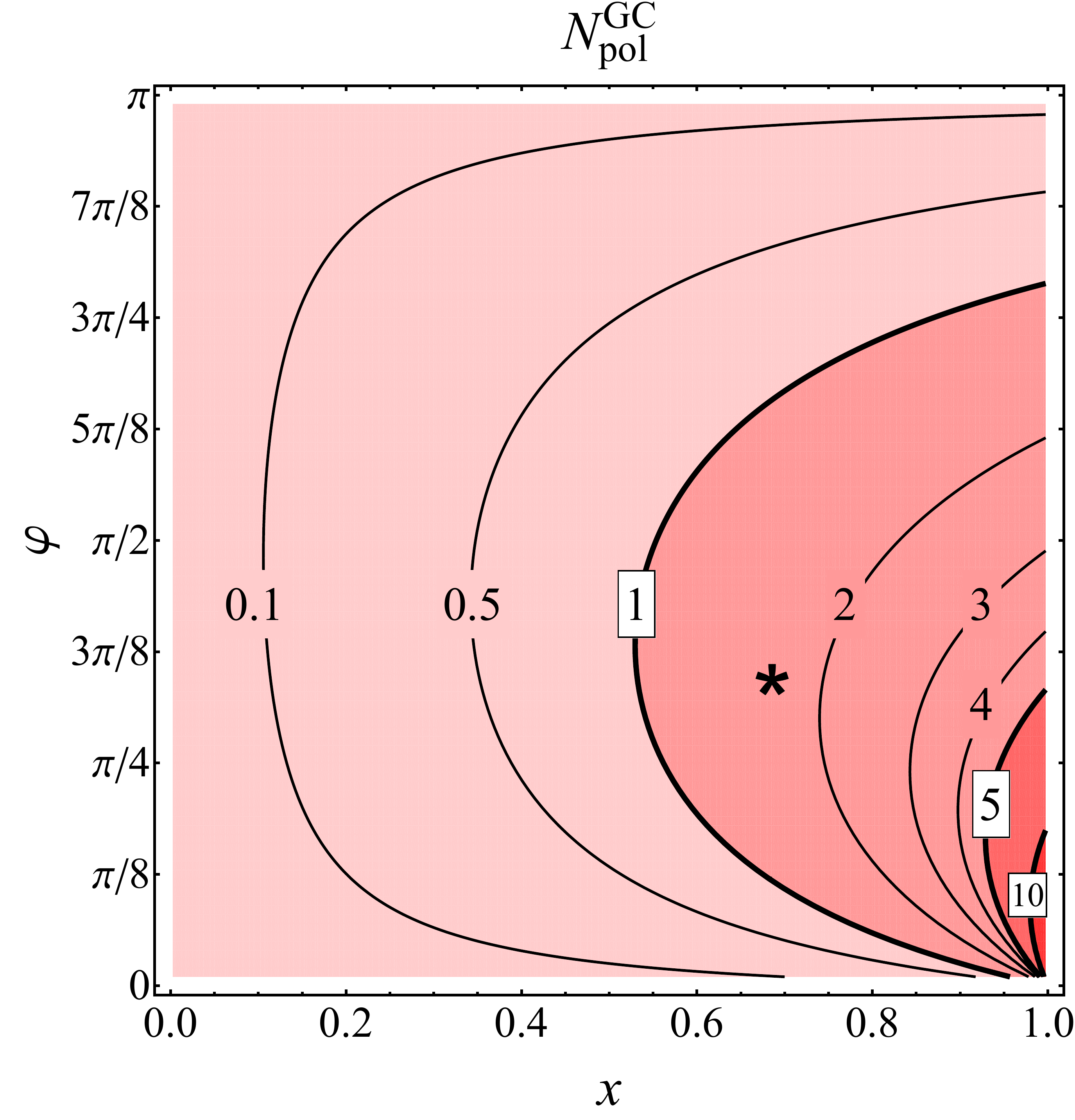}
\includegraphics*[width=0.49\textwidth]{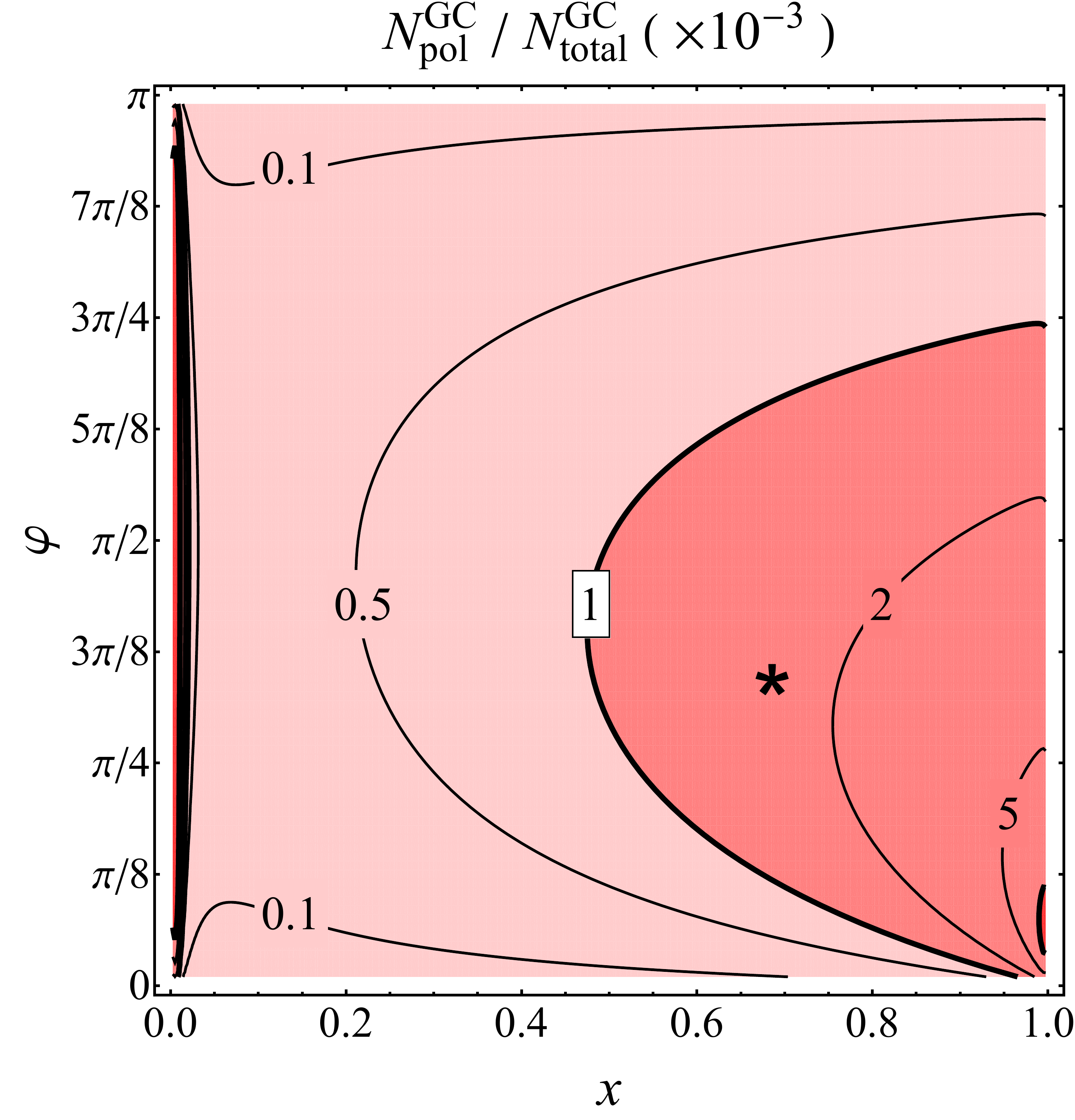}
\caption{\label{fig:SigEvents_SoverB} Contour plots in the $(x, \varphi)$-plane
of the $N_{\rm pol}^{\rm GC}$ (left panel) and $N_{\rm pol}^{\rm GC} / N_{\rm total}^{\rm GC}$ (right panel),
assuming at each point that
$|\lambda|^2$ is chosen to have the maximum value consistent with
observational constraints. The plot range is $0.002\leq x\leq 0.998$ and $0.01\pi\leq\varphi\leq 0.99\pi$. These two plots are made with the typical ACT parameters given at the beginning of Sec.~\ref{sec:ACTdetector}. The star denotes the benchmark point described in Table~\ref{tab:bench}.}
\end{figure}

We start with the symmetric dark matter case, using the ACT detector parameters given at the beginning of Sec.~\ref{sec:ACTdetector}.
In Figure~\ref{fig:SigEvents_SoverB}, we present contour plots of $N_{\rm pol}^{\rm GC}$ (left panel)
and $P_\gamma \equiv N_{\rm pol}^{\rm GC} / N_{\rm total}^{\rm GC}$ (right panel)  in the $(x, \varphi)$-plane, assuming that
at every point in the plane $|\lambda|^2$ is chosen to have the maximum value consistent with
observational constraints of INTEGRAL\footnote{In the right panel of Figure~\ref{fig:SigEvents_SoverB}, it seems that there is a rise in $P_\gamma$ close to $x=0$. This is due to the exponential cut-off in the background events as in Eq.~\eqref{eq:GCbackground}. Thus this effect is artificial since Eq.~\eqref{eq:GCbackground} is no longer valid for energy well above the cut-off.}.  Given our assumption of detector parameters
($I_{\rm exp} e_p = 0.5~\cm^2 ~\yr$ and $A=0.1$), it is clear that there is no point in the parameter space, except for the corner of $x\rightarrow 1$ and $\varphi\rightarrow 0$,
where the asymmetry in secondary particles, $AN_{\rm pol}^{\rm GC}$, would be even as large as one event.  For most of the parameter space, one would need
$I_{\rm exp} e_p$ to be a factor ${\cal O}(10-100)$ larger than our assumption to obtain
an expected secondary particle asymmetry event, even from the Galactic Center.  Thus, there would be no advantage
to considering a smaller angular window than the $60^\circ \times 10^\circ$ box on which we
have focused.
Since $N_{\rm pol}^{\rm Draco}$ is
much smaller still, we can indeed ignore dwarf spheroidals for the remainder of the analysis.
We note that by using the scaling behavior $N_{\rm pol}^{\rm GC}\sim I_{\rm exp}e_p$ and $N_{\rm pol}^{\rm GC}/N_{\rm total}^{\rm GC}\sim\epsilon^{-1}$, we can easily move to other detector parameters.

Note also that $N_{\rm decay}^{\rm GC} / N_{\rm astro}^{\rm GC} \ll 1$ and
$P_\gamma < 10^{-2}$
throughout most of the parameter space.  Thus,
a detection of a net circular polarization is only practical if systematic uncertainties
are smaller than $1\%$.  We have assumed that there are no systematic uncertainties arising from
astrophysical backgrounds, but there may be detector-related systematic uncertainties which would
need to be reduced below the $1\%$ level.   For the remainder of this analysis, we will assume
that systematic uncertainties can be reduced below this level.  As we have seen, using the Gaussian statistics, the signal significance can then be
determined as
\begin{equation}
\text{significance}=\frac{A N_{\rm pol}^{\rm GC}}{\sqrt{N_{\rm total}^{\rm GC}}} = A P_\gamma \sqrt{N_{\rm total}^{\rm GC}}\,,
\end{equation}
which scales with the detector parameters as $A (I_{\rm exp} e_p / \epsilon)^{1/2}$.  The signal-to-background
ratio $A \times  N_{\rm pol}^{\rm GC} / N_{\rm total}^{\rm GC} =AP_\gamma $, however, scales as $A \epsilon^{-1}$.

\begin{figure}[t]
\includegraphics[width=0.5\textwidth]{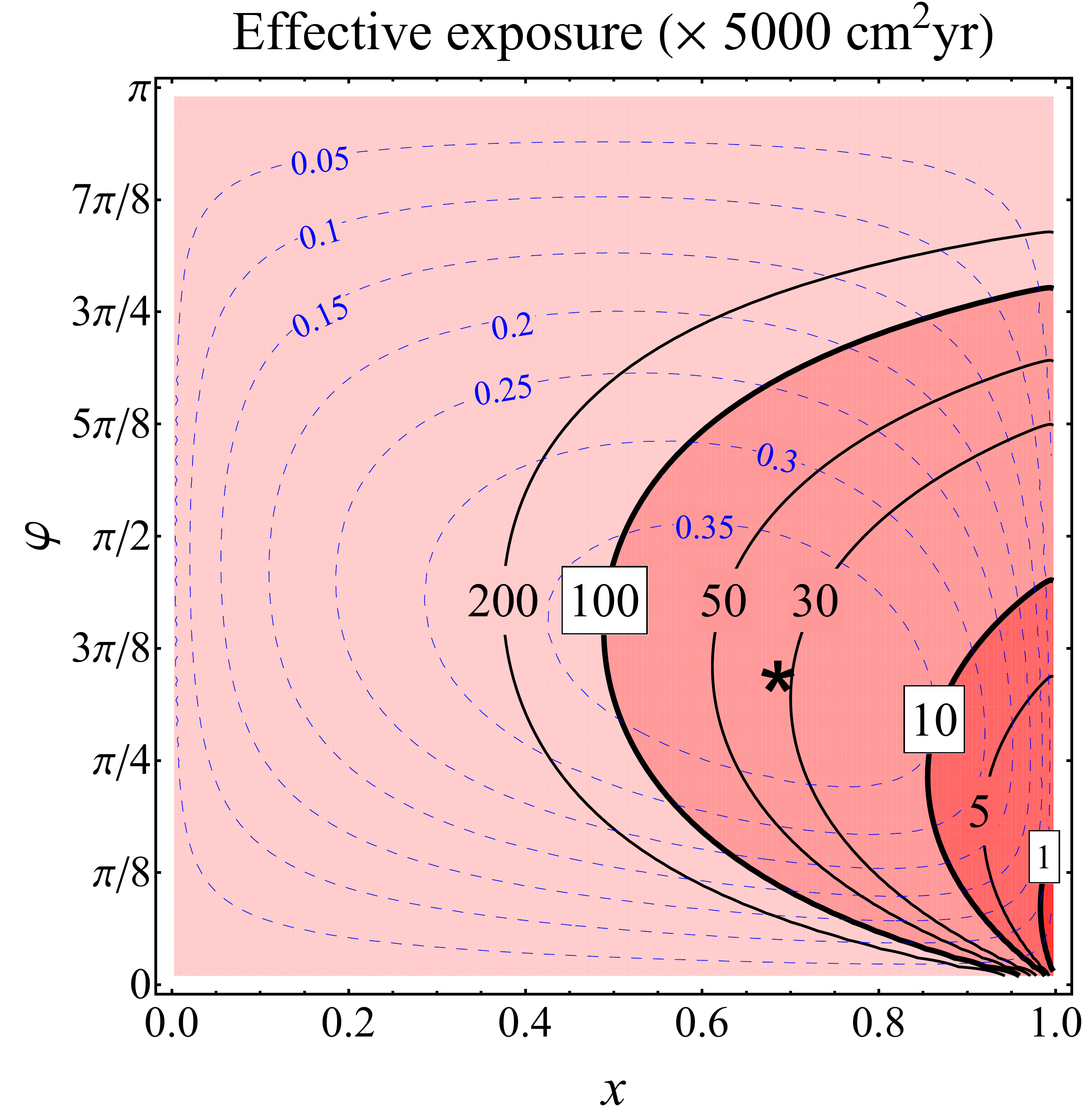}
\caption{\label{fig:exposure} Contour plots in the $(x, \varphi)$-plane
of the effective exposure $A^2I_{\rm exp} e_p/\epsilon$ (black solid contours) necessary
for $3\sigma$ evidence of a net circular polarization, assuming at each point that
$|\lambda|^2$ is chosen to have the maximum value consistent with
observational constraints.  The blue dashed lines are contours of $R(x, \varphi)$, and
the star denotes the benchmark point described in Table~\ref{tab:bench}.}
\end{figure}

In Figure~\ref{fig:exposure}, we plot contours of the effective exposure $A^2 I_{\rm exp} e_p / \epsilon$ needed to obtain
$3\sigma$ evidence for a net circular polarization in photons arising from the Galactic Center.  Contours are plotted in the $(x, \varphi)$-plane, assuming that
at every point in the plane $|\lambda|^2$ is chosen to have the maximum value consistent with
observational constraints.  In particular, we see that if $I_{\rm exp} = 5000~\cm^2~\yr$, then
one would need $A^2 e_p /\epsilon \gtrsim 30$ in order to find $3\sigma$ evidence of a net circular polarization at our benchmark point.

Note that the statistical significance of the monoenergetic line signal itself will be  enhanced by a factor
$(A^2 R^2 e_p)^{-1/2} >10^3$, and thus will be much larger than that of the
net circular polarization.  However, systematic uncertainties in the magnitude of the astrophysical foreground/background flux
are currently at the $2\%$ level, and by far dominate the statistical uncertainties.  Given the large number of photons
arriving from the direction of the Galactic Center (absent the efficiencies required for a circular polarization measurement),
even a line signal with an ${\cal O}(10^3)$ statistical significance would be obscured by systematic uncertainties.

Finally, we note the impact on this analysis of any weakening of the INTEGRAL bounds arising from
astrophysical uncertainties.  Any weakening of INTEGRAL's upper bound $\Gamma (X \rightarrow e^+ e^-)$ would
permit one to correspondingly increase $|\lambda|^2$, thus increasing $N_{\rm pol}^{\rm GC}$, but leaving
$N_{\rm astro}^{\rm GC} \sim N_{\rm total}^{\rm GC}$ unchanged.  The net circular polarization signal-to-background ratio and
signal significance thus both scale linearly with the INTEGRAL bound on $\Gamma(X \rightarrow e^+ e^-)$.

\subsubsection{The \texorpdfstring{$x \rightarrow 1$}{x->1} limit}

We end this subsection by discussing the interesting corner of parameter space in which $x \rightarrow 1$, $\varphi \rightarrow 0$.  In
this limit, $\Gamma (X \rightarrow \gamma \gamma) \rightarrow \alpha_{\rm em}^2 |\lambda|^2 m_e / 8\pi^3$ and
$R(x, \varphi) \rightarrow \pi \varphi \sqrt{1-x} \rightarrow 0$, but
\bea
\frac{\Gamma (X \rightarrow e^+ e^-)}{\Gamma (X \rightarrow \gamma \gamma)}
&\rightarrow& \frac{2\pi^2}{\alpha_{\rm em}^2} (1-x)^{1/2} \left[ (1-x) + \frac{\varphi^2}{4} \right] .
\eea
In this corner, the suppression of $\Gamma (X \rightarrow e^+ e^-)$ is more severe than the
suppression of $R$, so one can increase $N_{\rm pol}^{\rm GC}$  by increasing $|\lambda|$ and in turn $\Gamma (X \rightarrow \gamma \gamma)$,
without running afoul of INTEGRAL constraints on $\Gamma (X \rightarrow e^+ e^-)$.
However, this corner is cut off when $\lambda$ becomes so large that $\Gamma (X \rightarrow \gamma \gamma)$ itself
is constrained by the INTEGRAL bound, since the photons have $E_\gamma \sim m_e$ as $x \rightarrow 1$.

In this limit, the most optimistic sensitivity arises in the case where the
INTEGRAL line comes entirely from dark matter decay, with a negligible contribution
from any other astrophysical positron sources.   In this case, $N_{\rm total}^{\rm GC} \approx N_{\rm decay}^{\rm GC}$, and
$P_\gamma \approx R(x, \varphi)$.
Demanding
$\Gamma (X \rightarrow e^+ e^-) < \Gamma (X \rightarrow \gamma \gamma)$, we find the range of $x$ and $\varphi$ should be such that
\begin{equation}
(1-x)^{1/2}\left[(1-x)+\frac{\varphi^2}{4}\right]<\frac{\alpha_{\rm em}^2}{2\pi^2}\,,
\end{equation}
in which the value of $R(x,\varphi)\approx\pi\varphi(1-x)^{1/2}$ satisfies
\begin{equation}
    R(x,\varphi)\lesssim\left(\frac{\alpha_{\rm em}^2}{\sqrt{\pi}}\right)^{2/3}\frac{\sqrt{3}}{2}=8.4\times 10^{-4}\,.
\end{equation}
This region is roughly characterized by $(1-x)\lesssim (1/4)(\alpha^{2}_{\rm em}/\pi^2)^{2/3}=8\times 10^{-5}$ and $\varphi\lesssim\sqrt{3}(\alpha^2_{\rm em}/\pi^2)^{1/3}=0.03$. The total decay rate $\Gamma (X \rightarrow \gamma \gamma)$ then saturates the INTEGRAL
signal if $|\lambda| \sim {\cal O}(10^{-21})$.  The signal-to-background ratio is then
$\sim A \times N_{\rm pol}^{\rm GC} / N_{\rm decay}^{\rm GC} = A \times R(x, \varphi) < {\cal O}(10^{-4})$, implying that
systematic uncertainties must be controlled at a very
high level. We then find that the signal significance is given by
\bea
A R (x, \varphi) \sqrt{N_{\rm decay}^{\rm GC}} &\approx& 41 R (x, \varphi) \left( \frac{A^2 I_{\rm exp} e_p }{0.005~\cm^2~\yr} \right)^{1/2}
\lesssim  35 \left( \frac{A^2 I_{\rm exp} e_p}{5000~\cm^2~\yr} \right)^{1/2} .
\eea
Note that the signal-to-background ratio $A \times R(x, \varphi)$ is independent of
the INTEGRAL bound on $\Gamma(X \rightarrow \gamma \gamma)$ in the $x \rightarrow 1$ limit, while the
signal significance scales as $[\Gamma(X \rightarrow \gamma \gamma)]^{1/2}$.

\subsection{Detection Prospects for a Search of the Galactic Center for Asymmetric Dark Matter Decay }

For the case of asymmetric dark matter, the analysis is again much more straightforward.  In the most
optimistic case, one can set $R =1$, and the photons arising from dark matter decay have
a 100\% circular polarization.  Moreover, since the only constraint on the overall decay rate is
that at most 2\% of the photon flux in any one energy bin (for an experiment with an energy
resolution similar to COMPTEL) be due to dark matter decay, we may
optimistically set $P_\gamma = 4 \times 10^{-4} / \epsilon$.  The signal significance
may then be expressed as
\bea
\text{significance} &\approx& A P_\gamma \sqrt{N_{\rm astro}^{\rm GC}}
= 9.7 \left( \frac{A^2 I_{\rm exp} e_p / \epsilon}{5000~\cm^2~\yr} \right)^{1/2} ,
\eea
for $E_\gamma \sim \mev$.
One would thus need an effective exposure of $A^2 I_{\rm exp} e_p / \epsilon \sim 500~\cm^2~\yr$, in the most
optimistic scenario, to obtain $3\sigma$ evidence for a net photon circular polarization arising from
the decay of asymmetric dark matter.
This is an order of magnitude smaller than the effective
exposure required for $3\sigma$ evidence of the most optimistic scenario for symmetric dark matter.

\section{Conclusions \label{sec:Conclusions}}

We have considered scenarios in which interactions of dark matter with Standard Model
particles can result in the production within astrophysical objects of monoenergetic photons with a
net circular polarization.  Our focus has been on assessing the possibility of detecting such a
net circular polarization with future instruments.  Essentially, our goal has been to find the ``brightest
theoretical lamp posts" under which one could search for such a net circular polarization, and then assess
the type of instrument which would be needed to find evidence under these lamp posts.  Our particular analysis
has been focused on two possibilities, symmetric dark matter and asymmetric dark matter.

Interestingly, we find an almost unique scenario for symmetric dark matter
which presents the best prospect for detection.  In this scenario,
hyperstable spin-0 dark matter with mass $\gtrsim \mev$ has a $CP$-violating coupling to electrons, and can decay
to a monoenergetic photon pair through a one-loop process.  The choice of scenario is dictated by two requirements:
a circular polarization asymmetry can only arise if the mediators themselves can be produced through a tree-level decay
process, and the energy of the photons must be less than $\sim 30 \mev$ in order for them to be detected through
Compton scattering, which is sensitive to the photon polarization.  One finds that the net circular polarization can be as
large as $\sim 40\%$, and the photons will be in an energy range for which the detection of net circular polarization
through Compton scattering may be a viable detection strategy.

But unfortunately, these requirements both serve to hamper detection prospects, even for this optimal scenario.
The problem is that one is forced into a model in which the dark matter can also decay to a low-energy $e^+ e^-$ pair
at tree-level, and the rate for this process is constrained by data from INTEGRAL.

The optimal astrophysical target in which to search for such a signal would be the Galactic Center, but even for this
target, obtaining any evidence for a net circular polarization (for most of the parameter space) would require:
\begin{itemize}
\item{the instrument to have $(A^2 I_{\textbf{exp}} e_p / \epsilon) \gtrsim 50000~\cm^2~\yr$.  }
\item{the systematic uncertainties in the detector to be controlled at the $< 1\%$ level.}
\end{itemize}

But detection prospects would be improved in the case of asymmetric dark matter. In this case,
many of the theoretical constraints on the scenario disappear because $CP$-violation is no longer required in the interactions,
since the initial state itself breaks $CP$.  Indeed, it is easy to arrange a scenario where an asymmetric dark matter
particle decays only to a daughter particle $\psi$ and a photon ($\chi \rightarrow \psi \gamma$)
at tree-level with 100\%
polarization.  In this case, $3\sigma$ evidence of the circular polarization
could be found for $(A^2 I_{exp} e_p / \epsilon)$ as small as $500~\cm^2~\yr$.

We have seen that it would be challenging to detect a net circular
photon polarization arising from dark matter decay, even under the optimal versions of this scenario.
Given $A \sim 0.1$, $e_p \sim 10^{-4}$ and $\epsilon \sim 0.01$, which are reasonable assumptions given
the current state of detector technology, one would need an exposure of at least  $I_{\rm{exp}} \gtrsim  \mathcal{O}(10^7~\cm^2 ~\yr)$ to obtain
evidence for circular polarization arising from the decay of asymmetric dark matter.  For symmetric dark
matter, this minimum exposure required throughout most of the parameter space
is $I_{\rm{exp}} \gtrsim \mathcal{O}(10^9~\cm^2 ~\yr)$.
But for either case, however, it seems that new
technology might be required.

It would be interesting to further explore new detection technologies which could be used in detecting
a net circular polarization in photons, especially at higher energies.  If such an analysis could be done
for photons in the ${\cal O}(100)\mev$ range, then one could study models in which symmetric dark matter
decayed to net circularly polarized photons through a loop of muons, instead of electrons.  In this case,
there would be no bound on the total decay rate from INTEGRAL, and the allowed regions of
parameter space would be more amenable to detection.

But even in the asymmetric dark matter scenario, for which there is no
bound from INTEGRAL, detection prospects are still challenging due to the $A^2 e_p \sim 10^{-6}$ suppression
of the effective area.   Perhaps the best prospects lie with increasing the efficiency for producing useful events ($e_p$) and
the secondary asymmetry ($A$).

\acknowledgments

We are grateful to Kimberly Boddy, Celine Boehm, Celine Degrande, Philip von Doetinchem, Tathagata Ghosh, Danny Marfatia, Olivier Mattelaer,
Kuver Sinha, Aaron Vincent and Lian-Tao Wang for useful discussions.
AE acknowledges support by DOE Grant DE-SC-0008172 and NSF Grant PHY-1066014.
JK is supported in part by NSF CAREER Grant PHY-1250573.
PS has been supported in part by NSF Grant No. PHY-1417367 and is now supported in part by NSF Grant No. PHY-1720282. FT is supported in part by the Knut and Alice Wallenberg Foundation under grant KAW 2013.0235 and the Ragnar S\"{o}derberg Foundation under grant S1/16.

\bibliography{Refs}

\appendix

\section{\label{sec:decayrate}Derivation of the Decay Rate to Two-photon Final State}
Before beginning the calculation, we first define the notations to be used in this Appendix. We denote the momentum and polarization of the two final state photons as $(p_1,\epsilon_1)$ and $(p_2,\epsilon_2)$. In the rest frame of the dark matter, the three-momenta of the photons satisfy $\pmb{p}_1+\pmb{p}_2=0$ and $|\pmb{p}_1|=|\pmb{p}_2|=(m_X/2)$. Following the convention of~\cite{Elvang:2013cua}, we gauge-fix the polarizations to the following forms:
\begin{align}
\label{eq:gauge}
    & (\epsilon_1^+)_{\alpha\dot\alpha}=\sqrt{2}\,\frac{|1]_{\alpha}\langle 2|_{\dot\alpha}}{\langle 21\rangle} & & (\epsilon_1^-)_{\alpha\dot\alpha}=\sqrt{2}\,\frac{|2]_{\alpha}\langle 1|_{\dot\alpha}}{[21]} \nonumber\\
    & (\epsilon_2^+)_{\alpha\dot\alpha}=\sqrt{2}\,\frac{|2]_{\alpha}\langle 1|_{\dot\alpha}}{\langle 12\rangle} & & (\epsilon_2^-)_{\alpha\dot\alpha}=\sqrt{2}\,\frac{|1]_{\alpha}\langle 2|_{\dot\alpha}}{[12]}\,,
\end{align}
where $\epsilon_{\alpha\dot\alpha}=(\epsilon\cdot\sigma)_{\alpha\dot\alpha}$ is the spinor helicity form~\cite{Xu:1986xb} of the polarization $\epsilon$. Under this gauge, we have
\begin{align}
    & \epsilon_{1}^{\pm}\cdot\epsilon_{2}^{\pm}=-1 & & \epsilon_{1}^{\pm}\cdot\epsilon_{2}^{\mp}=0 & & p_i\cdot\epsilon_j=0\quad\forall i,j\in\{1,2\}\,.
\end{align}
Finally, for our special kinematics, the nonzero spinor inner products between the two photons are $[12]=\langle 12\rangle=im_X$.

The Feynman diagrams (a) and (b) of Figure~\ref{fig:feyn} give us the loop integrand of $X\rightarrow\gamma\gamma$:
\begin{align}
\label{eq:Integrand}
    \mathcal{I}=|\lambda|e^2\cos\frac{\varphi}{2}\left(\frac{\Tr[\mathcal{O}_1]}{d_0d_1d_3}+\frac{\Tr[\mathcal{O}_2]}{d_0d_2d_3}\right)-i|\lambda|e^2\sin\frac{\varphi}{2}\left(\frac{\Tr[\mathcal{O}_1\gamma^5]}{d_0d_1d_3}+\frac{\Tr[\mathcal{O}_2\gamma^5]}{d_0d_2d_3}\right)\,.
\end{align}
In the numerators, $\mathcal{O}_{1,2}$ are the matrices
\begin{align}
    & \mathcal{O}_1=(\slashed{l}+m_e)\slashed{\epsilon}_1(\slashed{l}-\slashed{p}_1+m_e)\slashed{\epsilon}_2(\slashed{l}-\slashed{p}_1-\slashed{p}_2+m_e)\nonumber\\
    & \mathcal{O}_2=(\slashed{l}+m_e)\slashed{\epsilon}_2(\slashed{l}-\slashed{p}_2+m_e)\slashed{\epsilon}_1(\slashed{l}-\slashed{p}_1-\slashed{p}_2+m_e)
\end{align}
where $l$ is the loop momentum. In the denominators, we have
\begin{align}
    & d_0=l^2-m_e^2 & & d_1=(l-p_1)^2-m_e^2 \nonumber\\
    & d_2=(l-p_2)^2-m_e^2 & & d_3=(l-p_1-p_2)^2-m_e^2\,.
\end{align}
To proceed, we first work out the gamma matrix traces in the first term of Eq.~\eqref{eq:Integrand}:
\begin{align}
    & \Tr[\mathcal{O}_1]=4m_e\left[4(\epsilon_1\cdot l)(\epsilon_2\cdot l)+\frac{m_X^2}{2}+d_1\right]\nonumber\\
    & \Tr[\mathcal{O}_2]=4m_e\left[4(\epsilon_1\cdot l)(\epsilon_2\cdot l)+\frac{m_X^2}{2}+d_2\right]\,.
\end{align}
Then following~\cite{Ellis:2011cr}, the loop integral gives
\begin{equation}
    \int\frac{d^4l}{(2\pi)^4}\frac{\Tr[\mathcal{O}_1]}{d_0d_1d_3}=\int\frac{d^4l}{(2\pi)^4}\frac{\Tr[\mathcal{O}_2]}{d_0d_2d_3}=\frac{im_e}{2\pi^2}\left[(1-x)f(x)-1\right]\,,
\end{equation}
where $x=(2m_e/m_X)^2$ and $f(x)$ is the result of the triangle integral:
\begin{equation}
    f(x)=\frac{m_X^2}{2}\int\frac{d^4l}{i\pi^2}\frac{1}{d_0d_1d_3}=\left\{\begin{array}{ll}
    -\left[\arcsin(x^{-1/2})\right]^2 & x>1 \\
    \frac{1}{4}\left[\log\frac{1+\sqrt{1-x}}{1-\sqrt{1-x}}-i\pi\right]^2 & x\leqslant 1
    \end{array}\right.\,.
\end{equation}
Therefore, the first term of Eq.~\eqref{eq:Integrand} gives
\begin{equation}
\label{eq:I1}
    |\lambda|e^2\cos\frac{\varphi}{2}\int\frac{d^4l}{(2\pi)^4}\left(\frac{\Tr[\mathcal{O}_1]}{d_0d_1d_3}+\frac{\Tr[\mathcal{O}_2]}{d_0d_2d_3}\right)=\frac{2i\alpha_{\text{em}}|\lambda|m_e}{\pi}\times\cos\frac{\varphi}{2}\left[(1-x)f(x)-1\right]\,.
\end{equation}
This term is independent of the polarizations of the final state photons.

Next, we calculate the second term of Eq.~\eqref{eq:Integrand}. The gamma matrix traces in the numerators can be simplified to
\begin{align}
    \Tr[\mathcal{O}_1\gamma^5]=\Tr[\mathcal{O}_2\gamma^5]&=\left(-4im_e\right)\varepsilon^{\mu\nu\rho\lambda}(\epsilon_1)_{\mu}(p_1)_{\nu}(\epsilon_2)_{\rho}(p_2)_{\lambda}\nonumber\\
    &=-m_e\left(\varepsilon_{\alpha\gamma}\varepsilon_{\beta\delta}\varepsilon_{\dot\alpha\dot\delta}\varepsilon_{\dot\beta\dot\gamma}-\varepsilon_{\alpha\delta}\varepsilon_{\beta\gamma}\varepsilon_{\dot\alpha\dot\gamma}\varepsilon_{\dot\beta\dot\delta}\right)(\epsilon_{1})^{\dot\alpha\alpha}(p_1)^{\dot\beta\beta}(\epsilon_2)^{\dot\gamma\gamma}(p_2)^{\dot\delta\delta}\,,
\end{align}
where we have used the spinor helicity representation of the antisymmetric tensor $\varepsilon^{\mu\nu\rho\lambda}$. Now plugging in $(\epsilon_1)$ and $(\epsilon_2)$ given in Eq.~\eqref{eq:gauge}, and
\begin{equation*}
(p_1)^{\dot\beta\beta}=|1\rangle^{\dot\beta}[1|^{\beta}\qquad (p_2)^{\dot\delta\delta}=|2\rangle^{\dot\delta}[2|^{\delta}\,,
\end{equation*}
we get:
\begin{equation}
    \Tr[\mathcal{O}_1\gamma^5]=\Tr[\mathcal{O}_2\gamma^5]=\left\{\begin{array}{l@{\hspace{2em}}l}
    -2m_em_X^2 & \gamma^+\gamma^+\text{ final state} \\
    2m_em_X^2 & \gamma^-\gamma^-\text{ final state}
    \end{array}\right.\,.
\end{equation}
Therefore, the integral of the second term of Eq.~\eqref{eq:Integrand} yields
\begin{equation}
\label{eq:I2}
    -i|\lambda|e^2\sin\frac{\varphi}{2}\int\frac{d^4l}{(2\pi)^4}\left(\frac{\Tr[\mathcal{O}_1\gamma^5]}{d_0d_1d_3}+\frac{\Tr[\mathcal{O}_2\gamma^5]}{d_0d_2d_3}\right)=-\frac{2\alpha_{\text{em}}|\lambda|m_e}{\pi}\times (\pm 1)\sin\frac{\varphi}{2}f(x)\,,
\end{equation}
where the $\pm$ sign corresponds to the $\gamma^{\pm}\gamma^{\pm}$ final state. By combining Eq.~\eqref{eq:I1} and \eqref{eq:I2}, we get our final result,
\begin{align}
    &\mathcal{A}_{\gamma^+\gamma^+}=\frac{2i\alpha_{\text{em}}|\lambda|m_e}{\pi}\left[\left[(1-x)f(x)-1\right]\cos\frac{\varphi}{2}+if(x)\sin\frac{\varphi}{2}\right]\nonumber\\
    &\mathcal{A}_{\gamma^-\gamma^-}=\frac{2i\alpha_{\text{em}}|\lambda|m_e}{\pi}\left[\left[(1-x)f(x)-1\right]\cos\frac{\varphi}{2}-if(x)\sin\frac{\varphi}{2}\right]\,,
\end{align}
which is nothing but Eq.~\eqref{eq:A2gamma}.

\section{Propagation of Statistical Uncertainties on Asymmetry Measurements}
\label{sec:appB}

An asymmetry can be measured by defining two different kinematics regions in the final state parameter space and counting particles in each of the two regions. We denote by $n_+$ and $n_-$ the number of particles in the first and the second regions. An asymmetry is then defined as:

\begin{equation}
\label{eq:B1}
F = \frac{n_+ - n_-}{n_+ + n_-}\,,
\end{equation}
The statistical uncertainty on $F$ is
\begin{equation}
\label{eq:B2}
\sigma_F = \sqrt{  \left(\frac{\partial F}{\partial n_+} \sigma_{n_+} \right)^2
+ \left(\frac{\partial F}{\partial n_-} \sigma_{n_-} \right)^2 +2\frac{\partial F}{\partial n_+}\frac{\partial F}{\partial n_-}cov(n_+,n_-) }\,,
\end{equation}
where
\begin{align}
&\frac{\partial F}{\partial n_+} = \frac{2n_-}{(n_+ + n_-)^2}\nonumber\\
&\frac{\partial F}{\partial n_-} = -\frac{2n_+}{(n_+ + n_-)^2}\,.
\end{align}
Considering $n_+$ and $n_-$ to be independently gaussian distributed, we have
\begin{align}
&\sigma_{n_+} = \sqrt{n_+} \,, \nonumber\\
&\sigma_{n_-} = \sqrt{n_-} \,, \nonumber\\
&cov(n_+,n_-) = 0\,.
\end{align}
Therefore,
\begin{equation}
\label{eq:B5}
\sigma_F = \frac{2}{n_+ + n_-}\sqrt{\frac{n_+n_-}{n_+ + n_-}}\,.
\end{equation}
Introducing $N=n_+ + n_-$ and noting that $n_+ = \frac{N(1+F)}{2}$ and  $n_- = \frac{N(1-F)}{2}$, using Eq.~\eqref{eq:B5}, we have
\begin{equation}
\label{eq:B6}
\sigma_F = \sqrt{\frac{1-F^2}{N}}\,.
\end{equation}
Consequently, to measure $F$ at the $3\sigma$ level, one has to satisfy $\frac{F}{\sigma_F} \geq 3$, and therefore
\begin{equation}
\label{eq:B7}
N \geq 9\frac{1-F^2}{F^2}\,.
\end{equation}

\end{document}